\documentclass[12pt]{article}
\usepackage{pst-plot,url,epsf}
\newcommand{\beq}{\begin{equation}}
\newcommand{\eeq}{\end{equation}}
\newcommand{\bea}{\begin{eqnarray}}
\newcommand{\eea}{\end{eqnarray}}
\newcommand{\beas}{\begin{eqnarray*}}
\newcommand{\eeas}{\end{eqnarray*}}

\newcommand{\epm}{e^+e^-}

\newcommand{\ra}{\rightarrow}

\newcommand{\udmn}{e^+ e^- \ra b u \bar{d} \;\bar b \mu^- \bar{\nu}_{\mu} \;
                     b \bar{b}}

\newcommand{\AmS}{{\protect\the\textfont2
  A\kern-.1667em\lower.5ex\hbox{M}\kern-.125emS}}

\newcommand{\nn}{\nonumber}

\begin{document}
\thispagestyle{empty}
\begin{flushright}
December 2006\\
Revised:\\
 May 2007\\
\vspace*{1.5cm}
\end{flushright}
\begin{center}
{\LARGE\bf Off mass shell effects in associated
           production of the top quark pair and Higgs boson
           at a linear collider\footnote{Work supported in part by
           the Polish State Committee for Scientific Research in years 
           2006--2008}}\\
\vspace*{2cm}
Karol Ko\l odziej\footnote{E-mail: kolodzie@us.edu.pl} 
and Szymon Szczypi\'nski\footnote{E-mail: simon@server.phys.us.edu.pl}\\[6mm]
{\small\it Institute of Physics, University of Silesia\\ 
ul. Uniwersytecka 4, PL-40007 Katowice, Poland}\\
\vspace*{2.5cm}
{\bf Abstract}\\
\end{center}
We discuss effects related to the fact that the final state particles
of the reaction $\epm \ra t \bar t H$ are actually produced and they decay
off mass shell. 
For the intermediate mass Higgs boson, 
which decays preferably into a $b\bar b$-quark pair, the
reaction will be observed through
reactions with 8 fermions in the final state. Such reactions,
already in the lowest order of the standard model, receive contributions
typically from a few dozen thousands of the Feynman diagrams, 
the vast majority of which constitute background
to the signal of associated production of the top quark pair and Higgs 
boson. In order to illustrate pure off mass shell effects we neglect the background
contributions and compare the `signal' cross section 
with the cross section in the narrow width approximation for
$\udmn$, which is one of possible detection channels of 
the associated production of the top quark pair and Higgs boson
at a linear collider.

\vfill

\newpage

\section{INTRODUCTION}

If the Higgs boson exists in Nature then it will be most probably discovered
at the Large Hadron Collider, but the accurate study of its production and decay
properties, which is crucial for verification of electroweak (EW) symmetry breaking 
mechanism, can be best performed in a clean environment of $\epm$ collisions at 
the future International Linear Collider (ILC) \cite{ILC}. The study
would be of the utmost importance for establishment of the Standard Model (SM)
or possibly some of its extensions as, {\em e.g.} the minimal supersymmetric SM 
(MSSM) or some of more general supersymmetric models
as the theory of the EW interactions. 
The Higgs boson mass $m_H$ can be constrained in the framework of SM
by the virtual effects it has on precision EW observables.
The combined value of the top quark mass measured at Tevatron and the 
combined $W$-boson mass \cite{Hmass} give a central value
of $m_H=85_{-28}^{+39}$ GeV and an upper limit of 166~GeV, both at
95\% CL, in agreement with
combined results on the direct searches for the Higgs boson at LEP that
lead to a lower limit of $114.4$~GeV at 95\% CL \cite{LEPdir}.
These constraints indicate that the SM Higgs boson should be
searched for in the mass range just above the lower direct search limit 
\cite{Kilminster}.

\begin{figure}[htb]
\vspace{140pt}
\centerline{
\includegraphics{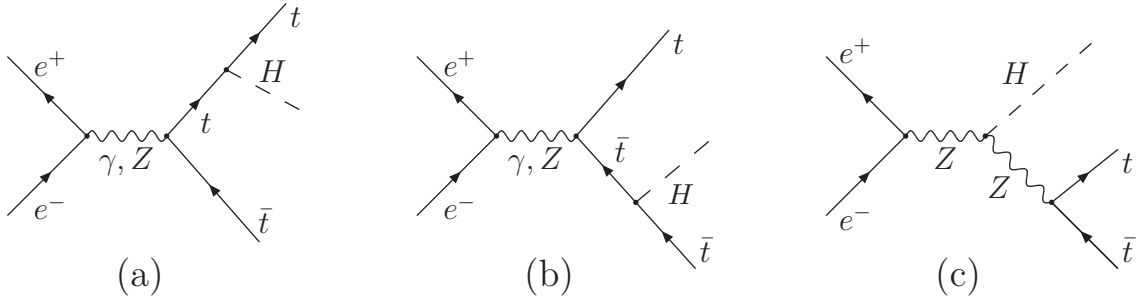}\hfill}
\caption{Feynman diagrams of reaction (\ref{eetth}) to the lowest order of SM
         with the neglect of the scalar boson couplings to electrons.}
\label{fig:eetth}
\end{figure}

If the Higgs boson mass is below the top quark pair threshold, $m_{H} < 2 m_{t}$, 
then the Higgs Yukawa coupling to the top quark $g_{ttH}$ can be directly determined 
in the process of associated production of the top quark pair and Higgs boson
\cite{eetth}
\bea
\label{eetth}
e^+ e^- \rightarrow t \bar{t} H.
\eea
The lowest order SM Feynman diagrams of reaction (\ref{eetth}), with the neglect
of the scalar boson couplings to electrons, 
are shown in Fig.~\ref{fig:eetth}.
As the contribution of the Higgs boson emission off the virtual $Z$-boson line, 
which is represented by the diagram in Fig.~\ref{fig:eetth}c,
is small with respect to the Higgsstrahlung off the top quark line illustrated
in Fig.~\ref{fig:eetth}a and \ref{fig:eetth}b, the SM lowest
order cross section of reaction (\ref{eetth}) becomes practically
proportional to $g_{ttH}^2$. This fact makes  
reaction (\ref{eetth}) so sensitive to the Higgs Yukawa coupling to the top quark.
Because of its numerical value close to 1, precise determination of $g_{ttH}$
may play an indispensable role in our understanding of the whole mass generation 
mechanism of SM.

Because of their large decay widths, the $t$- and $\bar{t}$-quark of reaction 
(\ref{eetth})
almost immediately decay into $bW^+$ and $\bar{b}W^-$, respectively, the $W$-bosons
subsequently decay into 2 fermions each and the Higgs boson, if it is lighter than
140 GeV, decays dominantly into a $b \bar b$-quark pair. Thus reaction (\ref{eetth})
will be actually detected at the ILC as a reaction of the form
\beq
\label{ee8f}
  e^+e^-\;\; \ra \;\; bf_1\bar{f'_1} \bar{b}f_2 \bar{f'_2} b \bar b,
\eeq
where $f_1, f'_2 =\nu_{e}, \nu_{\mu}, \nu_{\tau}, u, c$ and 
$f'_1, f_2 = e^-, \mu^-, \tau^-, d, s$. 
The three possible detection channels 
of (\ref{ee8f}), which
correspond to the decay modes of the $W$-bosons resulting from 
decays of the $t$- and $\bar t$-quark in the `signal' diagrams of reaction 
(\ref{eetth}), are 
\begin{enumerate}
\item hadronic channel: eight jets (38 \%),
\item  semileptonic channel: lepton and six jets  (37 \%),
\item leptonic channel: two leptons and four jets  (25 \%).
\end{enumerate}

In addition to the analysis in ref. \cite{eetth}, reaction (\ref{eetth}) 
has been extensively studied
in literature. The QCD radiative corrections to (\ref{eetth})
were calculated in \cite{QCDrcor}, $\mathcal{O}(\alpha)$ EW corrections were 
calculated in \cite{EWrcor} and full $\mathcal{O}(\alpha)$ EW and 
$\mathcal{O}(\alpha_s)$ QCD corrections were studied in \cite{Belanger}.
In \cite{Farrel}, process (\ref{eetth}) was considered
in the kinematic region where the Higgs boson energy is close to its
maximal energy and hence the next-to-leading-logarithmic corrections to
the Higgs boson energy distribution 
can be computed within the nonrelativistic effective theory.
Processes of the form $\epm \ra b\bar b b\bar b W^+W^- \ra b\bar b b\bar b
l^{\pm}\nu_lq\bar q'$ accounting for the signal of associated Higgs boson
and top quark pair production, as well as several irreducible background
reactions, were studied in \cite{Moretti} and EW contributions to
the leptonic and semileptonic reactions (\ref{ee8f}) have been 
computed in \cite{Schwinn}.
Moreover, feasibility of the measurement of the Higgs-top Yukawa coupling at the ILC 
in reaction (\ref{eetth}) was discussed in \cite{expfeas}.

Reactions (\ref{ee8f}), with 8 fermions in the final state, receive contributions
typically from a few dozen thousands of the Feynman diagrams already
in the lowest order of SM. Most of the diagrams constitute the 
`non signal' background to the reaction of the associated on shell 
top quark pair and Higgs boson production and their subsequent decay 
\beq
\label{eetth8f}
  e^+e^-\;\; \ra \;\; t \bar t H \;\; \ra \;\; 
             bf_1\bar{f'_1} \bar{b}f_2 \bar{f'_2} b \bar b,
\eeq
with the same final state as that of (\ref{ee8f}).

In the present note, we will look into pure off shell effects
that are related to the fact that $t$, $\bar t$ and $H$ of (\ref{eetth8f})
are actually produced and they decay as off mass shell particles.
To illustrate these effects, we will calculate the cross section of one
selected reaction (\ref{ee8f})
while keeping only the `signal' Feynman diagrams 
and compare it with the cross in the narrow width approximation (NWA)
for the top, antitop and Higgs in the energy range that could be 
relevant for the ILC. 
We will also check whether or not the off shell effects 
change the prediction of \cite{eetth} that the cross section of (\ref{eetth})
is dominated by the Higgs boson emission off $t$ and $\bar t$. We realize
that neglecting an overwhelming number of the `non signal' Feynman diagrams
may be too crude an approximation, in particular in the energy range not
far above the $t\bar t H$ threshold. However, our simplified approach has the 
advantage that it fully takes into account spin correlations that are
of great importance in the context of top quark physics \cite{Jezabek}.
Taking into account the spin correlations would in particular increase the sensitivity
to new physics effects, such as an anomalous $Wtb$ couplings
\cite{wtb}, especially
if the beam polarization is provided at the ILC.

\section{OUTLINE OF THE CALCULATION}

\begin{figure}[htb]
\vspace{140pt}
\includegraphics{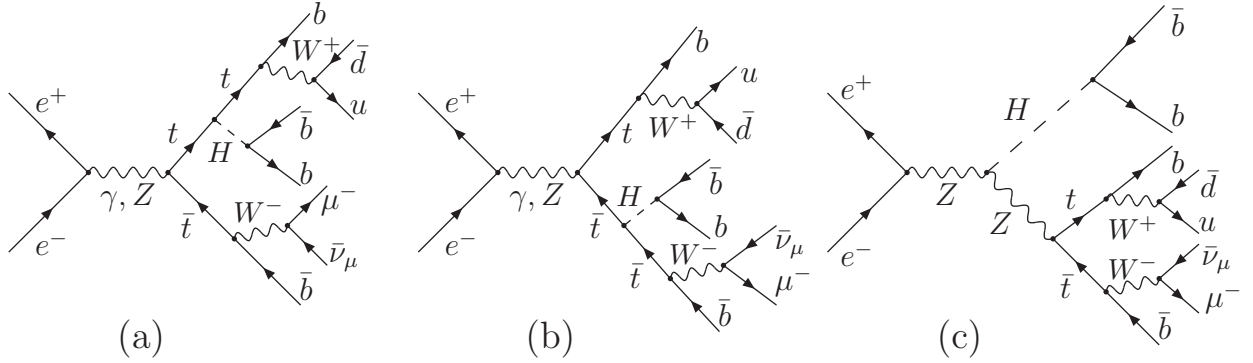}
\caption{`Signal' Feynman diagrams of reaction (\ref{udmn}).}
\label{fig:udmn}
\end{figure}

We will concentrate on one selected semileptonic channel of reaction (\ref{ee8f}), 
namely
\bea
\label{udmn}
e^+(p_1)\; e^-(p_2) \; \ra \; b(p_3) \; u(p_4)\; \bar{d}(p_5) \; \bar b(p_6) \;
\mu^-(p_7) \; \bar{\nu}_{\mu}(p_8) \; b(p_9) \; \bar{b}(p_{10}),
\eea
where the particle four momenta have been indicated in parentheses.
In the unitary gauge, reaction (\ref{udmn}) receives
contributions from 56550 Feynman diagrams already in the lowest order of SM. 
The corresponding amplitudes can be
generated with a Fortran 90 program {\tt carlomat} written by one of 
the present authors \cite{carlomat}. 
If we neglect the Higgs boson couplings 
to fermions lighter than a $b$-quark
then the number of diagrams is reduced to 26816. However,
as already stated in Section~1, in order to illustrate the size of the pure off 
mass shell effect we will restrict ourselves to the lowest order 
`signal' Feynman diagrams shown in Fig.~\ref{fig:udmn}. 
The corresponding
amplitudes are calculated with the helicity amplitude method that had
been described in detail in \cite{JKZ} and computed with the use of
program libraries described in \cite{programs}.

With that restricted numbers of amplitudes, the actual challenge of the computation
of the total cross section of (\ref{udmn}) is a right choice of parametrization 
of the 20-fold 
phase space integral which should lead to reliable results without using 
vast numbers of calls to the integrand.
The 3 different phase space parametrizations which we use are the following
\bea
\label{dps12}
 {\rm d}^{20} Lips  &=& (2\pi)^{-20}  
          {\rm d} PS_2\left(s,s_{345910},s_{678}\right)
          {\rm d} PS_2\left(s_{345910},s_{345},s_{910}\right) \nonumber \\
&\times&  {\rm d} PS_2\left(s_{345},m_3^2,s_{45}\right) 
          {\rm d} PS_2\left(s_{678},m_6^2,s_{78}\right) \nonumber \\
&\times&  {\rm d} PS_2\left(s_{45},m_4^2,m_5^2\right)
          {\rm d} PS_2\left(s_{78},m_7^2,m_8^2\right) 
          {\rm d} PS_2\left(s_{910},m_9^2,m_{10}^2\right) \nonumber \\
&\times&  {\rm d} s_{345910} {\rm d} s_{678} {\rm d} s_{345} {\rm d} s_{45} 
          {\rm d} s_{78} {\rm d} s_{910},\\
\label{dps34}
 {\rm d}^{20} Lips  &=& (2\pi)^{-20}  
          {\rm d} PS_2\left(s,s_{345},s_{678910}\right)
          {\rm d} PS_2\left(s_{345},m_3^2,s_{45}\right) \nonumber \\
&\times&  {\rm d} PS_2\left(s_{45},m_4^2,m_5^2\right)
          {\rm d} PS_2\left(s_{678910},s_{678},s_{910}\right) \nonumber \\
&\times&  {\rm d} PS_2\left(s_{678},m_6^2,s_{78}\right) 
          {\rm d} PS_2\left(s_{78},m_7^2,m_8^2\right) 
          {\rm d} PS_2\left(s_{910},m_9^2,m_{10}^2\right)\nonumber \\
&\times&  {\rm d} s_{345} {\rm d} s_{678910} {\rm d} s_{45} {\rm d} s_{678}
          {\rm d} s_{78} {\rm d} s_{910},\\
\label{dps5}
 {\rm d}^{20} Lips  &=& (2\pi)^{-20}  
          {\rm d} PS_2\left(s,s_{345678},s_{910}\right)
          {\rm d} PS_2\left(s_{345678},s_{345},s_{678}\right) \nonumber \\
&\times&  {\rm d} PS_2\left(s_{345},m_3^2,s_{45}\right) 
          {\rm d} PS_2\left(s_{45},m_4^2,m_5^2\right)
          {\rm d} PS_2\left(s_{678},m_6^2,s_{78}\right) \nonumber \\
&\times&  {\rm d} PS_2\left(s_{78},m_7^2,m_8^2\right) 
          {\rm d} PS_2\left(s_{910},m_9^2,m_{10}^2\right)\nonumber \\
&\times&  {\rm d} s_{345678} {\rm d} s_{345} {\rm d} s_{678} {\rm d} s_{45} 
          {\rm d} s_{78} {\rm d} s_{910}.
\eea
In Eqs.~(\ref{dps12})--(\ref{dps5}), $s=(p_1+p_2)^2$, 
$s_{ij...}=(p_i+p_j+...)^2,
i,j = 3,...,10$, and ${\rm d} PS_2\left(q^2,q_1^2,q_2^2\right)$
is a two particle (subsystem) phase space element defined by
\beq
 {\rm d} PS_2\left(q^2,q_1^2,q_2^2\right) = \delta^{(4)}\left( q - q_1 - q_2 \right) 
   \frac{{\rm d}^3q_1}{2E_1} \frac{{\rm d}^3q_2}{2E_2} 
    = \frac{|\vec{q}_1|}{4\sqrt{q^2}} {\rm d} \Omega_1,
\label{dps2}
\eeq
where $\vec{q}_1$ is the momentum and $\Omega_1$ is the solid angle
of one of the particles (subsystems) in the relative centre of mass system, 
$\vec{q}_1 + \vec{q}_2 = 0$. 

Parametrizations (\ref{dps12}), (\ref{dps34}) and (\ref{dps5}) have been chosen 
in such a way
that invariants $s_{ij...}$ correspond to virtualities of 
propagators of the gauge bosons, Higgs boson and/or top quarks in the diagrams
of Fig.~\ref{fig:udmn}. 
Possible poles in the propagators of unstable particles are regularized with 
the constant particle widths $\Gamma_a$  which are introduced through the complex 
mass parameters $M_a^2$ by making the substitution
\beq
\label{m2}
m_a^2 \;\ra \; M_a^2=m_a^2-im_a\Gamma_a, \qquad a=Z, W, H, t.
\eeq
In order to reduce variance of the MC integration invariants $s_{ij...}$ 
that are related to the resonating propagators of $W$, $H$ 
and $t$ are obtained from the random variables uniformly distributed in
the interval $[0,1]$ by performing
mappings smoothing out their Breit--Wigner distributions. Denote
the lower and upper physical limit of $s_{ij...}$ by $s_{ij...}^{\rm min}$ and
$s_{ij...}^{\rm max}$, respectively, and
the uniform random variable by $x \in [0,1]$, then the mapping is given
by
\bea
s_{ij...}=\Gamma_a m_a \tan\left(\frac{\Gamma_a m_a}{N} x + x_0\right)+m_a^2,
\label{BW}
\eea
with
\bea
N=\frac{\Gamma_a m_a}{\arctan\left(
\frac{s_{ij...}^{\rm max}-m_a^2}{\Gamma_a m_a}-x_0\right)} \qquad
{\rm and}
\qquad
x_0=-\arctan\left(\frac{m_a^2-s_{ij...}^{\rm min}}{\Gamma_a m_a}\right).
\eea
Mapping (\ref{BW}) is
not performed for $s_{345910}$, $s_{678910}$ and 
$s_{345678}$ in the propagators of $t$ in Fig.~\ref{fig:udmn}a,
$\bar t$ in Fig.~\ref{fig:udmn}b and $Z$ in Fig.~\ref{fig:udmn}c, respectively,
which are far away from resonance in the centre of mass system energy (CMS)
range considered. Those three invariants, as well as all the remaining 
integration variables of Eqs.~(\ref{dps12})--(\ref{dps2}), which we
symbolically denote by $y_j \in [a_j,b_j]$,
are obtained from the uniform random variables $x_j \in [0,1]$ 
with a simple linear transformation
\beq
           y_j=\left(b_j - a_j\right) x_j + a_j.
\eeq

Weights $w_i$, $i=1,2,3$, with which each of the 3 phase space parametrizations 
(\ref{dps12})--(\ref{dps5})
contributes to the total cross section are determined in the initial scanning run
performed with all the initial weights equal to 1/3.
They are calculated as the ratios
\beq
\label{wi}
          w_i=\bar{\sigma}_i/\sum_{j=1}^3 \bar{\sigma}_j, \qquad i=1,2,3,
\eeq
where $\bar{\sigma}_i$, $i=1,2,3$, denotes the cross section obtained 
in the initial scan with 
phase parametrization (\ref{dps12}), (\ref{dps34}), (\ref{dps5}), respectively.
The final result for the total cross section $\sigma$ of (\ref{udmn}) in the
`signal' approximation is then calculated as the weighted average
\beq
\label{sigma}
          \sigma=\sum_{j=1}^3 w_i\sigma_j
\eeq
with $\sigma_j$, $j=1,2,3$ being the cross section computed  
with phase parametrization (\ref{dps12}), (\ref{dps34}), (\ref{dps5}), respectively.

The cross section of (\ref{udmn}) in the NWA, 
is defined in the following way
\bea
\label{NWA}
\sigma_{\rm NWA}=
\sigma (e^+ e^- \rightarrow t \bar{t} H)\times 
\frac{\Gamma_{W^+\rightarrow u \bar{d}}}{\Gamma_{W}}\times 
\frac{\Gamma_{W^-\rightarrow \mu^- \bar{\nu}_{\mu}}}{\Gamma_{W}}
\times\frac{\Gamma_{H\rightarrow b \bar{b}}}{\Gamma_{H}},
\eea
where $\sigma (e^+ e^- \rightarrow t \bar{t} H)$ denotes the total
cross section of (\ref{eetth}) and we have assumed 
\begin{displaymath}
\frac{\Gamma_{t \rightarrow b W^+}}{\Gamma_{t}}=
\frac{\Gamma_{\bar t \rightarrow \bar b W^-}}{\Gamma_{t}}=1.
\end{displaymath}
The 5-dimensional numerical integration that is necessary in order to compute
$\sigma (e^+ e^- \rightarrow t \bar{t} H)$ does not require
the multi-channel MC approach and can be performed with the single
phase space parametrization given by
\bea
\label{dpsnw}
{\rm d}^{5} Lips  &=& (2\pi)^{-5}  
          {\rm d} PS_2\left(s,s_{t \bar t},m_H^2\right)
          {\rm d} PS_2\left(s_{t \bar t},m_t^2,m_t^2\right)
          {\rm d} s_{t \bar t},
\eea
where $s_{t \bar t}=\left(p_t + p_{\bar t}\right)^2$, with $p_t$ and $p_{\bar t}$
being the four momenta of the on shell $t$ and $\bar t$ in (\ref{eetth})
and the two particle phase space element ${\rm d} PS_2$ defined in (\ref{dps2}).

The NWA formula of Eq. (\ref{NWA}) is obtained from the signal cross section 
of reaction
(\ref{udmn}) by making the following substitution for the resonance factors
$D_a\left(q^2\right)$, $a=t,W$, corresponding to each of the resonating 
propagators of the top quark and $W$-boson in the Feynman diagrams 
of Fig.~\ref{fig:udmn}
\bea
\label{dq2}
D_a\left(q^2\right)=\left[\left(q^2 - m_a^2\right)^2 
                   + \left(m_a\Gamma_a\right)^2\right]^{-1} 
\ra \quad K_a\delta \left(q^2 - m_a^2\right),
\eea
with normalization factors $K_a=\int\limits_{-\infty}^{\infty} {\rm d}q^2
D_a\left(q^2\right) = \pi / \left(m_a \Gamma_a\right)$. It is just the Dirac
delta function on the right hand side of (\ref{dq2}) which causes the signal 
cross section of (\ref{udmn}) to take the factorized form of (\ref{NWA}).
Obviously, substitution (\ref{dq2}) makes sense only if the width of unstable 
particle $\Gamma_a$ is much smaller than its mass $m_a$. The relative error 
associated to it is usually estimated as being of 
$\mathcal{O}\left(\Gamma_a/m_a\right)$, as required by dimensional analysis.
With the values of the top quark and $W$-boson masses and widths given in 
Eqs. (\ref{vmass}), (\ref{fmass}) and (\ref{widths}) we obtain uncertainties
of 0.9\% and 2.5\% for each of the resonating top quark and $W$-boson 
propagators. Hence we may expect a discrepancy between the signal and NWA cross 
sections of the order of a few per cent. This expectations will be confirmed by
the actual numerical results which are given in the next section. For a more 
detailed discussion of limitations of the NWA the reader is referred 
to \cite{Kauer}.

\section{NUMERICAL RESULTS}

The numerical results presented in this section have been obtained with the
following set of initial physical parameters. We have chosen
the Fermi coupling and fine structure constant in the Thomson limit
\bea
\label{params1}
G_{\mu}=1.16639 \times 10^{-5}\;{\rm GeV}^{-2}, \qquad
\alpha_0=1/137.03599911,
\eea
as well as the $W$- and $Z$-boson masses
\bea
\label{vmass}
m_W=80.419\; {\rm GeV},\qquad m_Z=91.1882\; {\rm GeV},
\eea
as the EW input parameters.
The top quark mass and the external fermion masses of reaction (\ref{udmn}) 
are the following:
\bea
\label{fmass}
m_t=174.3\;{\rm GeV}, \quad m_b=4.8\;{\rm GeV},\quad m_u=5\;{\rm MeV},
\quad m_d=10\;{\rm MeV},\\
m_e=0.51099892\;{\rm MeV}, \quad m_{\mu}=105.6583\;{\rm MeV}. \qquad\qquad \nn
\eea
The value Higgs boson mass is assumed at $m_H=130$~GeV.
The widths of unstable particles that are introduced through substitution
(\ref{m2}) are calculated to the lowest order of SM. This results in
\bea
\label{widths}
\Gamma_t=1.531\;{\rm GeV}, \qquad \Gamma_W=2.048\;{\rm GeV},
\qquad \Gamma_H=8.065\;{\rm MeV},
\eea
for the top quark, $W$-boson and Higgs boson widths, respectively.
The actual value of the $Z$-boson width is not relevant for our calculation,
as the $Z$-boson propagator is far off its mass shell in all the diagrams
of Fig.~\ref{fig:udmn}.

\begin{figure}[ht]
\begin{center}
\setlength{\unitlength}{1mm}
\begin{picture}(35,35)(40,-50)
\rput(5.3,-6){\scalebox{0.6 0.6}{\epsfbox{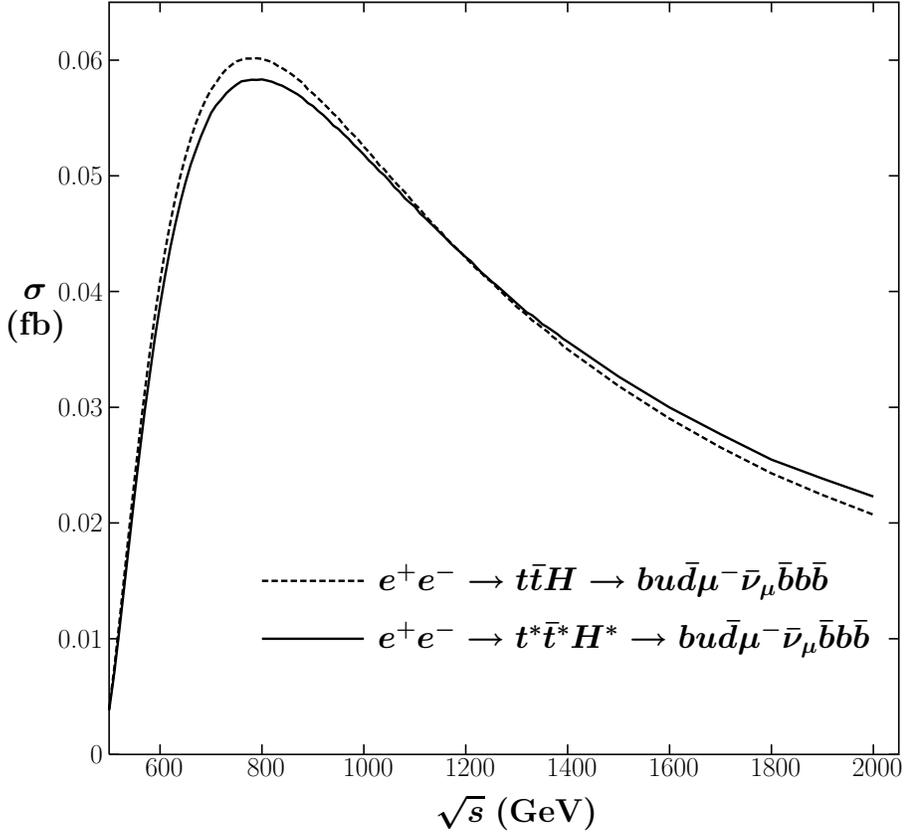}}}
\end{picture}
\end{center}
\vspace*{6.5cm}
\caption{Total cross sections of reaction (\ref{udmn}) as functions of the 
         CMS energy: the `signal' cross section $\sigma_{\rm signal}$ (solid line)
         and the cross section in NWA $\sigma_{\rm NWA}$ (dashed line).}
\label{fig:eetth_nwa.eps}
\end{figure}

The total cross section of reaction (\ref{udmn}) calculated with the `signal'
diagrams of Fig.~\ref{fig:udmn} is compared
with the cross section in the NWA
in Table~\ref{tab:8fnwa}. 
We see that the off mass shell effect is of the order of 3\%
for $\sqrt{s}=500-800$~GeV, {\em i.e.} close to the threshold of the associated top 
quark pair and Higgs boson production, then it decreases and becomes negative,
reaching an absolute value of about 5\% at $\sqrt{s}=2$~TeV.
This is illustrated in 
Fig.~\ref{fig:eetth_nwa.eps}, where both the `signal'
and NWA cross sections are plotted 
as functions of the CMS energy. 
\begin{table}
\begin{center}
\begin{tabular}{c|c|c|c}
\hline 
\hline 
\rule{0mm}{7mm} $\sqrt{s}$ [GeV]& $\sigma_{\rm signal}$ [ab] 
  & $\sigma_{\rm NWA}$ [ab] & $\delta$ [\%] \\[1.5mm]
\hline 
\hline 
\rule{0mm}{7mm}  500 & 3.805(11)  & 3.923(1)  &   3.1 \\ [1.5mm]
\hline 
\rule{0mm}{7mm}  800 & 58.33(7) & 60.07(3)    &   3.0  \\[1.5mm]
\hline 
\rule{0mm}{7mm} 1000 & 51.79(7) & 52.56(3)    &   1.5  \\[1.5mm]
\hline 
\rule{0mm}{7mm} 1200 & 42.99(6) & 42.96(3)    & $-0.1$ \\[1.5mm]
\hline 
\rule{0mm}{7mm} 2000 & 21.90(11)& 20.76(2)    & $-5.2$ \\[1.5mm]
\end{tabular} 
\end{center}
\caption{Total cross sections of reaction (\ref{udmn}): the `signal' cross
section $\sigma_{\rm signal}$, the cross section in NWA
$\sigma_{\rm NWA}$ and their relative difference
$\delta=\left(\sigma_{\rm NWA}-\sigma_{\rm signal}\right)
/\sigma_{\rm signal}$.
The numbers in parenthesis show the uncertainty of the last decimals.}
\label{tab:8fnwa}
\end{table}

The off shell effect is caused overwhelmingly by the nonzero widths of the top 
quarks 
and $W$-bosons, as for the assumed value of the Higgs boson mass, $m_H = 130$ GeV, 
the Higgs boson is very narrow, with a decay width of a few MeV.
Actually, the very small lowest order value of 8 MeV for $\Gamma_H$ 
in Eq.~(\ref{widths}) will be further substantially reduced if the QCD radiative 
corrections are taken into account, see, {\em e.g.} \cite{HDECAY} and \cite{JKW}.

It is interesting to see to which extent the off shell effects may
change the prediction of \cite{eetth} that the cross section of (\ref{eetth})
is dominated by the Higgs boson emission off the $t$- and $\bar t$-quark. 
To this end, in Table~\ref{tab:sigyuk}, we present 
the lowest order `signal' cross section of (\ref{udmn}) $\sigma_{\rm signal}$ and
the cross section $\sigma_{\rm signal}^{\rm no\; HZZ}$ that has been calculated 
with the diagrams of Fig.~\ref{fig:udmn}a and ~\ref{fig:udmn}b, {\em i.e.}
without the Higgsstrahlung off the 
$Z$-boson line represented by the diagram of Fig.~\ref{fig:udmn}c.
The corresponding results for the cross section in the NWA are given in 
Table~\ref{tab:sigyuk} as
$\sigma_{\rm NWA}$ and $\sigma_{\rm NWA}^{\rm no\; HZZ}$. Let us compare the 
relative differences shown in Table~\ref{tab:sigyuk} as
$\delta_1$ and $\delta_2$ defined by
\bea
\delta_1=\left(\sigma_{\rm signal}^{\rm no\; HZZ}-\sigma_{\rm signal}\right)/
\sigma_{\rm signal}, \qquad  \delta_2=\left(\sigma_{\rm NWA}^{\rm no\; HZZ}
-\sigma_{\rm NWA}\right)/\sigma_{\rm NWA}.
\label{deltas}
\eea
They quantify to which extent the Feynman diagram of 
Fig.~\ref{fig:udmn}c spoils proportionality of the cross section of 
(\ref{udmn}) to $g_{t\bar t H}^2$, which makes the measurement of the 
top-antitop-Higgs Yukawa coupling more difficult.
While $\delta_1$ takes into
account the fact that $t$, $\bar t$ and $H$ are produced and decay off mass shell, 
$\delta_2$ has been calculated assuming that they are on mass shell particles.
\begin{table}
\begin{center}
\begin{tabular}{c|c|c|c|c|c|c}
\hline 
\hline 
\rule{0mm}{7mm} $\sqrt{s}$ & $\sigma_{\rm signal}$ 
  & $\sigma_{\rm signal}^{\rm no\; HZZ}$ & $\delta_{1}$ 
                                               & $\sigma_{\rm NWA}$ 
  & $\sigma_{\rm NWA}^{\rm no\; HZZ}$ & $\delta_{2}$ \\[1.5mm]
    [GeV] & [ab]  & [ab] & [\%] & [ab] & [ab] & [\%] \\[1.5mm]
\hline 
\rule{0mm}{7mm}  500 & 3.805(11)  & 3.775(10)  &   $-0.8$ & 3.923(1)  & 3.886(1)  
                                                            &   $-0.9$ \\ [1.5mm]
\hline 
\rule{0mm}{7mm}  800 & 58.33(7) & 55.84(6)    &   $-4.3$ & 60.07(3) &  57.51(3)   
                                                            &   $-4.3$ \\[1.5mm]
\hline 
\rule{0mm}{7mm} 1000 & 51.79(7) & 48.52(6)    &   $-6.3$ & 52.56(3) &   49.23(3)  
                                                            &   $-6.3$ \\[1.5mm]
\hline 
\rule{0mm}{7mm} 1200 & 42.99(6) & 39.50(7)    & $-8.1$ & 42.96(3) &   39.41(3)  
                                                            & $-8.3$\\[1.5mm]
\hline 
\rule{0mm}{7mm} 2000 & 21.90(11)& 19.18(12)    & $-12.4$ & 20.76(2)&  18.03(2)   
                                                           & $-13.2$ \\[1.5mm]
\end{tabular} 
\end{center}
\caption{Total cross sections of reaction (\ref{udmn}): 
the `signal' cross
section $\sigma_{\rm signal}$, the `signal' cross
section calculated without the diagram of Fig.~\ref{fig:udmn}c 
$\sigma_{\rm signal}^{\rm no\;ZZH}$, 
the cross section in the NWA $\sigma_{\rm NWA}$, the cross
section in the NWA calculated without the diagram of Fig.~\ref{fig:udmn}c 
$\sigma_{\rm NWA}^{\rm no\;ZZH}$ and the relative differences
$\delta_1$ and $\delta_2$ of (\ref{deltas}).
The numbers in parenthesis show the uncertainty of the last decimals.}
\label{tab:sigyuk}
\end{table}
The lowest order `signal' cross section of (\ref{udmn}) $\sigma_{\rm signal}$
and the `signal' cross section without the diagram of Fig.~\ref{fig:udmn}c 
\begin{figure}[!ht]
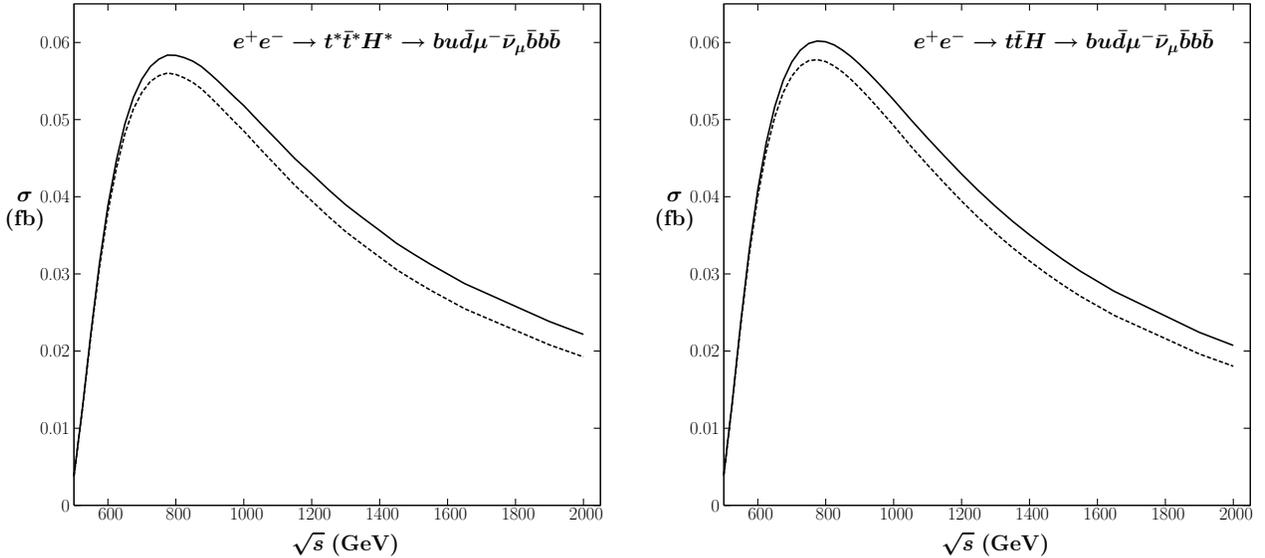

\begin{center}
\setlength{\unitlength}{1mm}
\begin{picture}(35,35)(60,-52)
\rput(5.3,-6){\scalebox{0.4 0.4}{\epsfbox{fig4a.epsi}}}
\end{picture}
\begin{picture}(35,35)(10,-52)
\rput(5.3,-6){\scalebox{0.4 0.4}{\epsfbox{fig4b.epsi}}}
\end{picture}
\end{center}
\vspace*{4.0cm}
\caption{The lowest order cross sections of (\ref{udmn}) as functions of 
         the CMS energy. On the left hand side: the `signal' cross section 
         (solid line) and the `signal' cross section without the diagram 
         of Fig.~\ref{fig:udmn}c (dashed line); on the right hand side:
         the corresponding cross sections in the NWA.}
\label{fig:sigyuk}
\end{figure}
$\sigma_{\rm signal}^{\rm no\; HZZ}$ are plotted as functions of the CMS energy
on the left hand side of Fig.~\ref{fig:sigyuk}. The plots on the right hand side
of Fig.~\ref{fig:sigyuk} show the corresponding cross sections in the NWA.
Both from Table~\ref{tab:sigyuk} and Fig.~\ref{fig:sigyuk}, we see that the 
Higgsstrahlung off the $Z$-boson line spoils the proportionality of the total 
cross section of the associated $t\bar t H$ production to $g_{t\bar t H}^2$ 
almost in the same way, independently of
whether the off shell effects are taken into account or not.

\section{Summary and outlook}

We have looked at a role that the off shell effects may play in  reaction 
(\ref{eetth}) of the associated production of the top quark pair and Higgs boson 
at the ILC. We have illustrated these effects for a semileptonic reaction
(\ref{udmn}), which is one of the detection channels of (\ref{eetth}) at the ILC,
by comparing the `signal' cross section that has been calculated by performing 
20-fold integration of the squared matrix element 
while keeping only the `signal' Feynman diagrams with the 
cross section in the NWA.
The off shell effects are typically of the order of a few per cent
for the CMS energies  in the range from 500 GeV to 2 TeV, in accordance
with the expectation based on the discussion of the quality of the NWA
in the end of Section 2.
We have also shown that the off shell effects do not affect much
the prediction of \cite{eetth} that the cross section of (\ref{eetth})
is dominated by the Higgs boson emission off $t$ and $\bar t$, which makes
it an attractive tool for determination of the Higgs--top Yukawa coupling.

The presented approach is very simplified, as a lot 
of the `non signal' background Feynman diagrams have been neglected in the
calculation of the lowest order cross section of (\ref{udmn}), but it has the 
advantage that it fully takes into account spin correlations that are
of great importance in the context of top quark physics.
Further work is needed in order to take into account 
a complete set of the lowest Feynman diagrams and include
leading radiative corrections for reactions (\ref{ee8f}).


\begin{thebibliography}{99}
\bibitem{ILC} J.A. Aguilar-Saavedra {\it et al.} [ECFA/DESY LC Physics
               Working Group Collaboration], arXiv:hep-ph/0106315;\\
               T.~Abe {\it et al.}, [American Linear Collider Working Group
               Collaboration],
               arXiv:hep-ex/0106056;\\
               K.~Abe {\it et al.} [ACFA Linear Collider Working Group
               Collaboration], arXiv:hep-ph/0109166.
\bibitem{Hmass} Ch. Parkes, International Conference on High Energy Physics,
                {\tt http://lephiggs.web.cern.ch/LEPEWWG}, July 2006.
\bibitem{LEPdir} R. Barate {\em et al.}, Phys. Lett. {\bf B565} (2003) 61.
\bibitem{Kilminster} B. Kilminster, arXiv:hep-ex/0611001, to appear in the 
                     proceedings of 33rd International Conference on High Energy 
                     Physics (ICHEP 06), Moscow, Russia, 26 Jul. -- 2 Aug. 2006. 
\bibitem{eetth} A. Djouadi, J. Kalinowski, P.M. Zerwas, Mod. Phys. Lett. {\bf A7}
                (1992) 1765;\\
                A. Djouadi, J. Kalinowski, P.M. Zerwas, Z. Phys {\bf C54} (1992) 255.
\bibitem{QCDrcor} S. Dittmaier, M. Kramer, Y. Liao, M. Spira, P.M. Zerwas, 
                  Phys. Lett. {\bf B441} (1998) 383;\\
                  S. Dittmaier, M. Kramer, Y. Liao, M. Spira, P.M. Zerwas, 
                  Phys. Lett. {\bf B478} (2000) 247;\\
                  S. Dawson, L. Reina, Phys Rev. {\bf D57} (1998) 5851;\\
                  S. Dawson, L. Reina, Phys Rev. {\bf D59} (1999) 054012.
\bibitem{EWrcor} Yu You {\it et al.}, Phys. Lett. {\bf B571} (2003) 85;\\
                 A. Denner, S. Dittmaier, M. Roth, M.M. Weber, Phys. Lett. {\bf B575}
                 (2003) 290. 
\bibitem{Belanger} G. B\'elanger {\it et al.}, Phys. Lett. {\bf B571} (2003) 163.
\bibitem{Farrel} C. Farrel, A.H. Hoang, Phys. Rev. {\bf D72}(2005) 014007.
\bibitem{Moretti} S. Moretti, Phys. Lett. {\bf B452} (1999) 338. 
\bibitem{Schwinn} C. Schwinn, arXiv:hep-ph/0412028.
\bibitem{expfeas} H. Baer, S. Dawson, L. Reina, Phys Rev. {\bf D61} (2000) 013002;\\
                  A. Juste, G. Merino, arXiv:hep-ph/9910301;\\
                  A. Juste, ECONF C0508141:ALCPG0426, 2005, arXiv:hep-ph/0512246;\\
                  A. Gay, arXiv:hep-ph/0604034.
\bibitem{Jezabek} M. Je\.zabek, J.H. K\"uhn, Nucl.Phys. {\bf B320}, 20 (1989).
\bibitem{wtb} B. Grz\c adkowski, Z. Hioki, Phys. Lett. {\bf B476}, 87 (2000);
              Phys. Lett. {\bf B529}, 82 (2002);
              Phys. Lett. {\bf B557}, 55 (2003);\\
              S.D. Rindani, Pramana {\bf 54}, 791 (2000), hep-ph/0002006;\\
              K. Ko\l odziej, 
              Phys. Lett. {\bf B584} (2004) 89. 
\bibitem{carlomat} K. Ko\l odziej, {\em ``{\tt carlomat}, a program for generation
                   of lowest order amplitudes''}, in preparation.
\bibitem{JKZ} K.~Ko\l odziej, M.~Zra\l ek, Phys.\ Rev.\ {\bf D43} (1991) 3619;\\
              F.~Jegerlehner, K.~Ko\l odziej, Eur.\ Phys.\ J.\ {\bf C12} (2000) 77.
\bibitem{programs} K.~Ko\l odziej, Comput. Phys. Commun. {\bf 151} (2003) 339;\\
                   K. Ko\l odziej, F. Jegerlehner, Comput. Phys. Commun. {\bf 159} 
                   (2004) 106.      
\bibitem{Kauer} N. Kauer, arXiv:hep-ph/0703077.
\bibitem{HDECAY} A. Djouadi, J. Kalinowski, M. Spira, Comput. Phys. Commun.
                  {108} (1998) 56.
\bibitem{JKW} F. Jegerlehner, K. Ko\l odziej, T. Westwa\'nski, 
              Eur. Phys. J. {\bf C 44} (2005) 195.
\end{thebibliography}
\end{document}